\documentclass[review]{elsarticle}
\usepackage{lmodern}
\usepackage{amsmath}
\usepackage{ifxetex,ifluatex}
\usepackage{amssymb}
\usepackage{bbm}
\usepackage{setspace}
\usepackage{geometry}
\usepackage{xcolor}
\newcommand{\rev}[1]{\textcolor{black}{#1}}

\usepackage{hyperref}

\journal{XXXXXX}

\begin{document}

\begin{frontmatter}

\title{Modeling repeated self-reported outcome data: a continuous-time longitudinal Item Response Theory model}

\author{Cécile Proust-Lima\corref{mycorrespondingauthor}}
\address{Univ. Bordeaux, Inserm, Bordeaux Population Health Research
Center, UMR1219, F-33000 Bordeaux, France}
\cortext[mycorrespondingauthor]{Corresponding author}
\ead{cecile.proust-lima@inserm.fr}

\author{Viviane Philipps}
\address{Univ. Bordeaux, Inserm, Bordeaux Population Health Research
Center, UMR1219, F-33000 Bordeaux, France}

\author{Bastien Perrot}
\address{Université de Nantes, Université de Tours, INSERM, SPHERE U1246, Nantes,
France}
\address{Methodology and Biostatistics Unit, CHU Nantes, Nantes, France}

\author{Myriam Blanchin}
\address{Université de Nantes, Université de Tours, INSERM, SPHERE U1246, Nantes, France}

\author{Véronique Sébille}
\address{Université de Nantes, Université de Tours, INSERM, SPHERE U1246, Nantes,
France}
\address{Methodology and Biostatistics Unit, CHU Nantes, Nantes, France}

\begin{abstract}
Item Response Theory (IRT) models have received
growing interest in health science for analyzing latent constructs such
as depression, anxiety, quality of life or cognitive functioning from
the information provided by each individual's items responses. However,
in the presence of repeated item measures, IRT methods usually assume
that the measurement occasions are made at the exact same time for all
patients. In this paper, we show how the IRT methodology can be combined
with the mixed model theory to provide a \rev{longitudinal} IRT model which exploits
the information of a measurement scale provided at the item level while
simultaneously handling observation times that may vary across
individuals \rev{and items}. The latent construct is a latent process defined in
continuous time that is linked to the observed item responses through a
measurement model at each individual- and occasion-specific observation
time; we focus here on a Graded Response Model for binary and ordinal
items. The Maximum Likelihood Estimation procedure of the model is available in the R package lcmm. The proposed approach is
contextualized in a clinical example in end-stage renal disease, the
PREDIALA study. The objective is to study the trajectories of depressive
symptomatology (as measured by 7 items of the Hospital Anxiety and
Depression scale) according to the time \rev{fom registration on the renal transplant waiting list}
and the renal replacement therapy. We also illustrate how the method can
be used to assess Differential Item Functioning and lack of measurement
invariance over time.
\end{abstract}


\end{frontmatter}

\section*{Highlights}

\begin{itemize}
\item
  The \rev{longitudinal} IRT model provides a flexible solution to analyze repeated
  item responses measuring a latent construct over time
\item
  The \rev{longitudinal} IRT model relies on a mixed model to capture the
  continuous-time nature of the underlying construct
\item
  The \rev{longitudinal} IRT model can investigate the item psychometric properties
  and measurement invariance
\item
  \rev{Maximum likelihood estimation} is made available in the R package
  lcmm with a companion vignette
\item
  The case study describes the depressive symptomatology of patients
  with end-stage renal disease on the transplant waiting list
\end{itemize}

\newpage
\section{Introduction}\label{introduction}

Item Response Theory (IRT) models, which exploit the information
provided by each individual's items responses, have received growing
interest in health science for capturing latent constructs of interest
such as depression, anxiety, fatigue, quality of life, or cognitive
functioning \cite{cerou2019,gorter2015,mccall2021,abdelhamid2021}. IRT models have
interesting properties compared to models coming from classical
measurement theory such as Classical Test Theory (CTT) models which
aggregate the items into a global score or score per domain. In
particular, CTT produces ordinal measurements while IRT generates
interval measurements. Hence, with IRT, a unit difference characterizes
the same amount when measured from different initial levels on the
latent construct scale. IRT also allows for \rev{analysis at the item-level} which enables a better
understanding of the item psychometric properties, and an in-depth
description of patients' experience.

In health research, the interest often lies in the longitudinal changes
of latent constructs. Examples include the trajectory of
anxiety or fatigue in clinical research \cite{rakers2021,otto2021} or the trajectory of functional dependency in epidemiological research on aging \cite{edjolo2016} based on the observed patients'
repeated responses to questionnaires either self-reported (and named
Patient-Reported Outcomes (PRO)) or reported by the clinician (and named
Clinician-Reported Outcomes (CRO)). IRT models have been extended \rev{to the modeling of latent constructs that change over time by leveraging repeated item measurements. However, most of them were built as repeated cross-sectional IRT models, and thus imposed that measurement occasions were occurring at the
exact same time for all patients \cite{wang2020,cai2021}}. This is, however,
rarely the case in practice. For instance, in cohort studies, even
though visits are planned, the exact timing may differ substantially
across individuals. The timescale may also be the time from a specific
health event (e.g.~diagnosis, registration on a waiting list),
independent from the time of study inclusion, so that the timescale
becomes \emph{per se} continuous. This is the case in aging studies where
trajectories are assessed according to age, or more generally in all the
situations where entry in the study does not correspond to a clearly
defined time zero. Models from linear mixed model theory \cite{laird1982} are particularly suited for the analysis of outcomes repeatedly measured over time. They can model the outcome trajectory in continuous
time while accounting for the within-subject correlation. We describe in
this paper how IRT modeling can be combined with linear mixed model
theory for the analysis of item responses measured repeatedly over time
when observation times vary across individuals. We show how to
operationalize the latent construct as a latent process defined in
continuous time and how to link it with the observed items responses
through a measurement model for graded responses, the Graded Response
Model (GRM) \cite{samejima1997}, at each individual- and occasion-specific
observation time. Beyond the description of a latent construct
trajectory over time from repeated item measurements, this \rev{longitudinal} IRT
model may also help to assess the item and scale properties as done with
cross-sectional IRT methods, and investigate lack of measurement
invariance, that is Differential Item Functioning (DIF) \rev{across} groups,
or Response Shift (RS) over more than two measurement times.

The proposed approach is contextualized in a clinical example in
end-stage renal disease, the PREDIALA study \cite{sebille2016,auneau-enjalbert2020}. The PREDIALA study aims at studying the
experience of patients with end-stage renal disease (e.g.~quality of
life, anxious and depressive symptoms) on the renal transplant waiting
list. From a psychological perspective, the waiting list period can be
long and anxiety-provoking for patients because of the uncertainty of
waiting, the hope of being called for a transplant and the
disappointment, and sometimes distress, of not being called. In
addition, the onset or worsening of depressive/anxious symptoms can
occur as the time on the waiting list increases \cite{tong2015}.
Moreover, patients' experience may also differ according to their renal
replacement therapy, that is, whether they are dialyzed or not (i.e.
preemptive) \cite{auneau-enjalbert2020}. As patients with chronic
diseases have to live and adapt to their illness, its stability or
progression over time, they may also understand or interpret the items
of the questionnaires differently according to socio-demographic or
clinical characteristics (e.g.~type of renal replacement therapy) or
over time, despite having similar health outcomes. The former situation
may induce DIF \cite{holland2009}, while the latter may produce
RS \cite{sprangers1999}. From a methodological perspective, the
relevant timescale was the time elapsed since registration on the
waiting list. Since entry in the study occurred after different periods
of time spent on the waiting list across individuals, this timescale
substantially differed from the classical follow-up time. It induces
large inter-individual variations of measurement times between patients
and calls for a \rev{longitudinal} IRT model that can handle individual-specific
measurement times. Such models can also enable the investigation of
measurement invariance \rev{across} groups (DIF) or over time (RS).

\hypertarget{methods}{%
\section{Methods}\label{methods}}

\hypertarget{dynamic-irt-model}{%
\subsection{\rev{Continuous-time longitudinal} IRT model}\label{dynamic-irt-model}}

In a sample of \(N\) subjects, let consider a set of \(K\) items
belonging to the same scale that are measured repeatedly over time, with
\(Y_{ikj}\) designating the response to item \(k\) (\(k=1,...,K\)) for
subject \(i\) (\(i=1,...,N\)) at repeated occasion \(j\)
(\(j=1,...,n_{ki}\)). Note that here, we consider the general framework
where the number of measurements may differ from one individual to the
other (and possibly from one item to the other) so that item measurement
\(Y_{ikj}\) is to be associated with its actual time of measurement
\(t_{ikj} \in \mathbb{R}\).

\hypertarget{structural-model}{%
\subsubsection{Structural model}\label{structural-model}}

As in classical IRT methodology, we assume that the \(K\) items measure
the same underlying construct called \(\Lambda\). The major difference
is that in a \rev{continuous-time} longitudinal IRT setting, the construct is now a latent
process \((\Lambda_i(t))_{t\in \mathbb{R}}\) defined in continuous time.

Its trajectory over time can be described by a linear mixed model (LMM)
to account for the within individual correlation (and between-individual
variability):

\begin{equation}\label{LMM}
\Lambda_i(t) = \boldsymbol{X}_{i}(t)^\top \boldsymbol{\beta} + \boldsymbol{Z}_{i}(t)^\top \boldsymbol{b}_i ~, ~~~~~~~ \forall t \in \mathbb{R}
\end{equation}

where \(\boldsymbol{X}_{i}(t)\) is a vector of variables including
functions of time \(t\), associated with parameters
\(\boldsymbol{\beta}\), which describes the shape of trajectory over time
of the construct of interest at the population level and its association
with covariates; \(\boldsymbol{Z}_i(t)\) is a vector of variables that
almost always includes exclusively functions of time and is associated
with the vector of individual random-effects \(\boldsymbol{b}_i\) with
\(\boldsymbol{b}_i \sim \mathcal{N}(\boldsymbol{0},\boldsymbol{B})\)
(\(\boldsymbol{B}\) is usually left unstructured). This second part
\(\boldsymbol{Z}_{i}(t)^\top \boldsymbol{b}_i\) models the individual departure from
the mean trajectory \(\boldsymbol{X}_{i}(t)^\top \boldsymbol{\beta}\). Although not considered in this work, adding a Gaussian stochastic process to structural mixed models may be of interest in some contexts to better reflect the local variations in the individual trajectories \citep{proust2006, proust-lima2013}. 

Identifiability constraint is added to this model to determine the
dimension of the latent process. Usually, the intercept is removed from
the model (i.e., \(\boldsymbol{X}_{i}(t)\) does not include any
intercept) so that the mean of the latent process is 0 in the category
of reference \rev{(that is for $\boldsymbol{X}_i(t)=\boldsymbol{0}$)}, and the variance of the first random-effect is constrained
to 1. This first random effect being usually an intercept, this
corresponds to assuming that the conditional variance given the
covariates is 1 at time \(t=0\).

\hypertarget{measurement-model}{%
\subsubsection{Measurement model}\label{measurement-model}}

The latent process is linked to the observations of the \(K\) items
using an item-specific measurement model. In this work, we focus mainly
on binary and ordinal items even though the methodology
could also apply to continuous items (see Proust-Lima et al. \cite{proust-lima2013} for more
details). We assume that item \(k\) is defined with \(L_k\) ordinal
levels from 0 to \(L_k-1\).

The probability to observe the level \(l\) for item \(Y_{ikj}\) is
defined by a cumulative probit model:

\begin{equation}\label{probit}
\mathbb{P}(Y_{ikj} \leq l ~|~\Lambda_{i}(t_{ikj})) = \Phi
\left ( \sigma^-_k ~ (\eta_{kl+1} - \Lambda_{i}(t_{ikj}))
\right )
\end{equation}

where \(\Phi\) is the Gaussian cumulative distribution function,
\(\sigma^-_k\) is a parameter defining the discrimination of item \(k\)
and \((\eta_{kl})_{l=1,...L_k-1}\) are the \(L_k-1\) location parameters
which correspond to the thresholds defining the change in the successive
levels of item \(k\). For an ordinal item, we assume
\(\rev{- \infty = \eta_{k0} \leq} \eta_{k1} \leq \eta_{k2} \leq ... \leq \eta_{kL_k-1}\leq \eta_{kL_k} = + \infty\).

Equation \eqref{probit} comes from the idea that item \(k\) takes the
level \(l\) if the underlying construct plus a measurement error
\(\epsilon_{ikj}\) of variance \(\sigma_k^2\) lies in the interval
\([\eta_{kl},\eta_{kl+1})\) as shown below:

\begin{equation}\label{explic}
\begin{array}{rcl}
\mathbb{P}(Y_{ikj} = l ~|~\Lambda_{i}(t_{ikj})) &=& \mathbb{P} \left ( \eta_{kl} \leq \Lambda_{i}(t_{ikj}) + \epsilon_{ikj} < \eta_{kl+1} \right ) \\
&=& \mathbb{P} \left ( \dfrac{\eta_{kl} - \Lambda_{i}(t_{ikj})}{\sigma_k} \leq  \dfrac{\epsilon_{ikj}}{\sigma_k} < \dfrac{\eta_{kl+1} - \Lambda_{i}(t_{ikj})}{\sigma_k} \right ) \\
&=& \mathbb{P} \left ( \dfrac{\epsilon_{ikj}}{\sigma_k} < \dfrac{\eta_{kl+1} - \Lambda_{i}(t_{ikj})}{\sigma_k}  \right ) - \mathbb{P} \left ( \dfrac{\epsilon_{ikj}}{\sigma_k} < \dfrac{\eta_{kl} - \Lambda_{i}(t_{ikj})}{\sigma_k} \right)
\end{array}
\end{equation}

By considering that the measurement error \(\epsilon_{ikj}\) is
Gaussian, it induces that

\begin{equation}\label{explic2}
\mathbb{P}(Y_{ikj} = l ~|~\Lambda_{i}(t_{ikj})) = \Phi \left( \dfrac{\eta_{kl+1} - \Lambda_{i}(t_{ikj})}{\sigma_k} \right ) - \Phi \left (\dfrac{\eta_{kl} - \Lambda_{i}(t_{ikj})}{\sigma_k} \right ) 
\end{equation}

and Equation \eqref{probit} comes by denoting
\(\sigma_k^- = \dfrac{1}{|\sigma_k|}\).

Note that, by considering logistic errors instead of Gaussian errors,
one would obtain the logistic ogive model, also very popular in IRT
methodology.

Equation \eqref{probit} defines a model for graded responses, also known
as Graded Response Model (GRM) \cite{samejima1997,baker2004}. We focus on this type of
model in the remaining of the work but acknowledge that any alternative
measurement model could be considered instead depending on the
distributional assumption and the items type (e.g., binary, continuous).
See for instance Saulnier et al. \citep{saulnier2021} in the same special issue and Barbieri et al. \citep{barbieri2017a}.

\hypertarget{measurement-invariance}{%
\subsubsection{Measurement invariance}\label{measurement-invariance}}

The \rev{longitudinal} IRT model defined with equations \eqref{LMM} and
\eqref{probit} assumes that all the items have the same functioning,
meaning that the common underlying construct captures all the
information of the items and what remains specific to the item is only
its location and discrimination/error; this is made clear with equation
\eqref{explic}. However, sometimes a different functioning of the items
may be suspected according to a covariate at a given time or over time.
In IRT methodology, this is called differential item functioning (DIF)
or Response Shift (RS) when lack of measurement invariance occurs over
time. DIF and RS can also be investigated in the \rev{longitudinal} IRT \rev{framework} by
completing the measurement model as follows:

\begin{equation}\label{probitDIF}
\mathbb{P}(Y_{ikj} \leq l ~|~\Lambda_{i}(t_{ikj})) = \Phi \left ( \sigma^-_k ~ (\eta_{kl} - (\Lambda_{i}(t_{ikj}) + (\boldsymbol{X}_i^{\text{DIF}}(t_{ikj}))^\top \boldsymbol{\gamma}_k)) \right )
\end{equation}

where \(\boldsymbol{X}_i^{\text{DIF}}(t)\) is the vector of covariates
for which a DIF is suspected and \(\boldsymbol{\gamma}_k\) the
associated parameters. When \(\boldsymbol{X}_i^{\text{DIF}}(t)\) is also
part of the structural model in Equation \eqref{LMM} (i.e,
\(\boldsymbol{X}_i^{\text{DIF}}(t) \subset \boldsymbol{X}_i(t)\)), an identifiability constraint is added to the \(\boldsymbol{\gamma}_k\)
parameters making them correspond to contrasts (i.e., deviations to the
mean effect) with \(\sum_{k=1}^{K} \boldsymbol{\gamma}_k =0\). The total
effect of \(\boldsymbol{X}_i^{\text{DIF}}(t)\) becomes the sum of its
common effect on the latent process (part of \(\boldsymbol{\beta}\)) and
its item-specific contrast \(\boldsymbol{\gamma}_k\).

With longitudinal data, two sorts of item-specific functionings may be
investigated:

\begin{itemize}
\item classical DIF with $\boldsymbol{X}_i^{\text{DIF}}(t)$ including time-independent covariates only; this explores how differently parameters of a specific item differ according to individual characteristics;

\item item response shift with $\boldsymbol{X}_i^{\text{DIF}}(t)$ including functions of time $t$; this explores how parameters of a specific item change over time. 
\end{itemize}

\hypertarget{maximum-likelihood-estimation}{%
\subsection{Maximum Likelihood
Estimation}\label{maximum-likelihood-estimation}}

We consider here a maximum likelihood framework for the estimation of
the \rev{longitudinal} IRT model.

\hypertarget{parameterization-of-the-vector-of-parameters}{%
\subsubsection{Parameterization of the vector of
parameters}\label{parameterization-of-the-vector-of-parameters}}

Let denote \(\boldsymbol{\theta}\) the total vector of parameters
defined in the structural part of the model described in \eqref{LMM} and
in the $K$ measurement equations described in \eqref{probit}. This vector
includes:

\begin{itemize}
\item the fixed effects $\boldsymbol{\beta}$ (except the intercept for identifiability).
\item the parameters specifying the Variance-Covariance matrix $\boldsymbol{B}$ of the random-effects. To ensure that $\boldsymbol{B}$ is positive definite, we consider the parameters of the Cholesky upper triangular transformation C (i.e., $\boldsymbol{B}=\boldsymbol{C}^\top \boldsymbol{C}$) with first element fixed at 1 for identifiability.
\item the discrimination parameters. We consider parameters $(\sigma_k)_{k=1,...K}$ so that the item discriminations $\sigma_k^- = \frac{1}{|\sigma_k|}$.
\item the vector of item $L_k - 1$ locations of each item $k$. To account for the constraint that $\eta_{k1} \leq \eta_{k2} \leq ... \leq \eta_{kL_k-1}$, we consider the vector $(\eta_{kl}^*)_{l=1,...,L_k-1}$  so that $\eta_{k1} = \eta_{k1}^*$ and $\eta_{kl} = \eta_{k1}^* + \sum_{m=2}^{l} \left ( \eta_{km}^* \right )^2$ for $l \geq 2$.
\item In case of differential effect of covariates on items (Equation \eqref{probitDIF}), the vector of parameters $(\boldsymbol{\gamma}_k)_{k=1,...,K-1}$ with $\boldsymbol{\gamma}_K = - \sum_{k=1}^{K-1} \boldsymbol{\gamma}_k$.
\end{itemize}

\hypertarget{contribution-to-the-likelihood}{%
\subsubsection{Contribution to the
likelihood}\label{contribution-to-the-likelihood}}

Let denote
\(\boldsymbol{\mathcal{Y}_i} = \left \{ Y_{ik}(t_{ikj}), \forall k \in \{1,...,K \}, \forall j \in \{1,...,n_{ik} \} \right \}\)
all the repeated item information of subject \(i\). The contribution of
subject \(i\) to the likelihood is

\begin{equation}\label{loglik}
\begin{array}{rcl}
l_i(\boldsymbol{\theta}) &=& \mathbb{P}(\boldsymbol{\mathcal{Y}}_i| \boldsymbol{\theta})  \\
&=& \int_{\mathbb{R}^p} \mathbb{P}(\boldsymbol{\mathcal{Y}}_i | \boldsymbol{\theta},\boldsymbol{b}_i) ~ f(\boldsymbol{b}_i) ~ d\boldsymbol{b}_i \\
&=& \int_{\mathbb{R}^p} \prod_{k=1}^K \prod_{j=1}^{n_{ik}} \prod_{l=0}^{L_{k-1}} \mathbb{P}(Y_{ik}(t_{ikj}) = l ~ |~ \boldsymbol{\theta},~ \boldsymbol{b}_i)^{\mathbbm{1}_{Y_{ik}(t_{ikj})=l}} ~ f(\boldsymbol{b}_i) ~ d\boldsymbol{b}_i \\
&=& \int_{\mathbb{R}^p} \prod_{k=1}^K \prod_{j=1}^{n_{ik}} \prod_{l=1}^{L_k}  \left \{ \Phi \left(  \sigma_k^- (\eta_{kl+1} - (\boldsymbol{X}_i(t_{ikj})^\top \boldsymbol{\beta} + \boldsymbol{Z}_i(t_{ikj})^\top \boldsymbol{b}_i))\right ) \right. \\
 & & ~~~~~~ - \left. \Phi \left ( \sigma_k^- (\eta_{kl} - (\boldsymbol{X}_i(t_{ikj})^\top \boldsymbol{\beta} + \boldsymbol{Z}_i(t_{ikj})^\top \boldsymbol{b}_i)) \right )  \right \} ~ f(\boldsymbol{b}_i) ~ d\boldsymbol{b}_i \\
\end{array}
\end{equation}

where the integration over the distribution of the $p$-vector of random
effects is obtained by Quasi Monte-Carlo approximation following
proposals of Philipson et al. \cite{philipson2020}. We systematically considered 1000
points in this work.

The maximum likelihood estimators of \(\boldsymbol{\theta}\) are
obtained by maximizing the log-likelihood
\(\mathcal{L} = \sum_{i=1}^N \log(l_i(\boldsymbol{\theta}))\). This is
achieved with the Marquardt-Levenberg algorithm, a robust Newton-like
algorithm, with stringent convergence criteria on the parameters, the
log-likelihood and the first and second derivatives of the
log-likelihood (see Philipps et al. \cite{philipps2021} for details).

The maximum likelihood estimates are denoted
\(\hat{\boldsymbol{\theta}}\) and their variance, obtained by the
inverse of the Hessian, is denoted
\(\widehat{V(\hat{\boldsymbol{\theta}})}\).

\hypertarget{software}{%
\subsubsection{Software}\label{software}}

The \rev{longitudinal} IRT model can be estimated with the \texttt{multlcmm}
function of R package \text{lcmm} \cite{proust-lima2017}. A package vignette provides a tutorial that fully describes the
present \rev{longitudinal} IRT model estimation and posterior computations on a
simulated dataset that mimics the PREDIALA data.

\hypertarget{posterior-computations}{%
\subsection{Posterior computations}\label{posterior-computations}}

We can compute several posterior quantities from the estimates
\(\hat{\boldsymbol{\theta}}\), and confidence intervals around these
quantities can be obtained by approximating the posterior distribution
by Monte-Carlo simulations using the asymptotic distribution of the
parameters
\(\boldsymbol{\theta} ~ \sim \mathcal{N}(\hat{\boldsymbol{\theta}},\widehat{V(\hat{\boldsymbol{\theta}})})\).
We consider for this 2000 random draws. In the following, we omit
\(\hat{\boldsymbol{\theta}}\) or more generally \(\boldsymbol{\theta}\)
in the equations for better readability.

\hypertarget{predicted-trajectories-of-the-construct}{%
\subsubsection{Predicted trajectories of the
construct}\label{predicted-trajectories-of-the-construct}}

Predicted trajectories of the construct can be computed either at the
population level (i.e., marginally to the random-effects) or at the
individual level (i.e., conditionally to the individual random-effects).
The predicted trajectory at the population level is computed for a
profile of covariates \(\boldsymbol{x}(t)\):

\[\mathbb{E}(\Lambda_i(t) | \boldsymbol{X}_i(t) = \boldsymbol{x}(t)) = \boldsymbol{x}(t)^\top \boldsymbol{\beta}\]
The predicted trajectory at the individual level is computed given the
individual covariates \(\boldsymbol{X}_i(t)\), \(\boldsymbol{Z}_i(t)\)
and all the information on the items
\(\boldsymbol{\mathcal{Y}}_i = \left \{ Y_{ik}(t_{ikj}), \forall k \in \{1,...,K \}, \forall j \in \{1,...,n_{ik} \} \right \}\):

\[\mathbb{E}(\Lambda_i(t) | \boldsymbol{X}_i(t),\boldsymbol{Z}_i(t),\boldsymbol{\mathcal{Y}}_i) = \boldsymbol{X}_i(t)^\top \boldsymbol{\beta} + \boldsymbol{Z}_i(t)^\top \mathbb{E}(\boldsymbol{b}_i | \boldsymbol{\mathcal{Y}}_i)\]
where the expected random-effect
\(\mathbb{E}(\boldsymbol{b}_i| \boldsymbol{\mathcal{Y}}_i)\) is
approximated by the mode of the posterior distribution
\(f(\boldsymbol{b}_i~ |~ \boldsymbol{\mathcal{Y}}_i)\).

\hypertarget{item-characteristic-curves}{%
\subsubsection{Item Characteristic
Curves}\label{item-characteristic-curves}}

With binary items, the Item Characteristic Curve (ICC) describes the
probability of the highest item level according to the underlying
construct level. With ordinal items, ICC translates into two curves:

\begin{itemize}
\item
  the item category probability curve also known as category
  characteristic curve which describes the probability of a response in
  a given item category according to the underlying construct level:

\begin{equation}\label{CCC}
\begin{array}{rcl}
\text{CCC}_{k,l}(\Lambda) &=& P(Y_{ik}(t)=l ~ | ~ \Lambda_i(t)=\Lambda) \\
&=& \Phi \left( {\sigma_k^-}~ (\eta_{kl+1} - \Lambda) \right ) - \Phi \left ({\sigma_k^-}~ (\eta_{kl} - \Lambda) \right ) 
\end{array}
\end{equation}

\item
  The item score expectation curve which can also be computed as a
  function of the underlying construct level:
\end{itemize}

\begin{equation}\label{ICC}
\begin{array}{rcl}
\mathbb{E}(Y_{ik} ~ | ~ \Lambda_i(t)=\Lambda) &=& \sum_{l=0}^{L_k-1} l ~ P(Y_{ik}(t)=l ~ | ~ \Lambda_i(t)=\Lambda) \\
&=& L_k-1 - \sum_{l=1}^{L_k-1} P(Y_{ik}(t) \leq l ~ | ~ \Lambda_i(t)=\Lambda) \\
&=& L_k-1 - \sum_{l=1}^{L_k-1} \Phi \left ( \sigma^-_k ~ (\eta_{kl} - \Lambda) \right ) 
\end{array}
\end{equation}

These two curves allow representing the items and their properties. The items locations describe where the items function along the construct level while the steepness of the item score expectation characterizes the items discriminations. For example, the steeper the
curve for an item, the better it can discriminate between two different
construct levels.

\hypertarget{predicted-trajectories-of-the-items}{%
\subsubsection{Predicted trajectories of the
items}\label{predicted-trajectories-of-the-items}}

The predicted trajectory over time of each item can be computed for a
profile of covariates \(\boldsymbol{x}(t)\) as follows:

\begin{equation}\label{predItem}
\mathbb{E} \left (Y_{ik}(t) | \boldsymbol{X}_i(t) = \boldsymbol{x}(t) \right )  = \int_{\mathbb{R}^p} \mathbb{E}(Y_{ik} (t) ~ | ~ \Lambda_i(t)= \boldsymbol{x}(t)^\top \boldsymbol{\beta} + \boldsymbol{z}(t)^\top \boldsymbol{b}) ~ f(\boldsymbol{b})~ d\boldsymbol{b}
\end{equation}

where
\(\mathbb{E}(Y_{ik} ~ | ~ \Lambda_i(t)= \boldsymbol{x}(t)^\top \boldsymbol{\beta} + \boldsymbol{z}(t)^\top \boldsymbol{b})\)
is computed as in equation \eqref{ICC} with
\(\Lambda = \boldsymbol{x}(t)^\top \boldsymbol{\beta} + \boldsymbol{z}(t)^\top \boldsymbol{b}\),
and the integral over the random-effects \(\boldsymbol{b}\) distribution is obtained
by Quasi Monte-Carlo approximation.

\hypertarget{fisher-information}{%
\subsubsection{Fisher Information}\label{fisher-information}}

The Fisher information provides a quantification of the level of
information brought by each item, and each item level. It is computed
using the second derivatives of the item level probability denoted
\(P_{kl}(\Lambda) = P(Y_{ik}(t)=l ~ | ~ \Lambda_i(t)=\Lambda)\) for item
\(k\) and level \(l\). The item information function for category \(l\)
is defined as follows (calculations are detailed in Section 1 of the
supplementary material):

\begin{equation}
\begin{split}
I_{kl}(\Lambda)P_{ikl}(\Lambda) &= - \dfrac{ \partial^2 \log \left ( P_{ikl}\left (\Lambda \right )\right )}{\partial \Lambda^2} P_{kl}\left (\Lambda \right )\\
&= - \dfrac{1}{\sigma_k^2} \dfrac{ \left (\phi \left (\dfrac{\Lambda - \eta_{kl+1}}{\sigma_ k}\right ) - \phi \left (\dfrac{\Lambda - \eta_{kl}}{\sigma_ k} \right )\right )^2}{\Phi \left (\dfrac{\Lambda - \eta_{kl+1}}{\sigma_ k}\right ) - \Phi \left (\dfrac{\Lambda - \eta_{kl}}{\sigma_ k}\right )}  \\
& ~~~~~~~~~ - \left ( \dfrac{\Lambda - \eta_{kl+1}}{\sigma_ k} \phi \left (\dfrac{\Lambda - \eta_{kl+1}}{\sigma_ k}\right ) - \dfrac{\Lambda - \eta_{kl}}{\sigma_ k} \phi \left (\dfrac{\Lambda - \eta_{kl}}{\sigma_ k}\right ) \right )
\end{split}
\end{equation}

The information curve provides the summary at the item level as follows:

\begin{equation}
\begin{split}
I_k(\Lambda) &= \sum_{l=1}^{L_k-1} I_{kl}(\Lambda)P_{kl}(\Lambda) \\
&= \sum_{l=1}^{L_k-1}  - \dfrac{1}{\sigma_k^2} \dfrac{ \left (\phi \left (\dfrac{\Lambda - \eta_{kl+1}}{\sigma_ k}\right ) - \phi \left (\dfrac{\Lambda - \eta_{kl}}{\sigma_ k} \right )\right )^2}{\Phi \left (\dfrac{\Lambda - \eta_{kl+1}}{\sigma_ k}\right ) - \Phi \left (\dfrac{\Lambda - \eta_{kl}}{\sigma_ k}\right )} 
\end{split}
\end{equation}

\hypertarget{application-to-prediala-study}{%
\section{Application to PREDIALA
study}\label{application-to-prediala-study}}

We applied the \rev{longitudinal} IRT model to analyze the repeated measures of
depressive symptomatology in the PREDIALA study. The HADS (Hospital
Anxiety and Depression scale) was used to measure anxiety and depression
disorders \cite{zigmond1983}. The HADS consists of 14 items on a
4-point Likert scale, seven of which are related to anxiety symptoms and
seven to depression symptoms. Only the depression symptom domain is
presented here. The 7 items rated from 0 (total agreement) to 3 (total
disagreement) are as follows:

\begin{itemize}
\item
  Item 2 ``I still enjoy the things I used to enjoy'' (Enjoy)
\item
  Item 4 ``I can laugh and see the funny side of things'' (Laugh)
\item
  Item 6 ``I feel cheerful'' (Cheerful)
\item
  Item 8 ``I feel I am slowed down'' (Slow)
\item
  Item 10 ``I have lost interest in my appearance'' (Appearance)
\item
  Item 12 ``I look forward with enjoyment to things'' (Looking forward)
\item
  Item 14 ``I can enjoy a good book or radio or TV programme'' (Leisure)
\end{itemize}

Responses to items 8 and 10 are reversed so that higher levels
systematically indicate more intense symptoms.

Our objective was to describe the trajectory of depressive
symptomatology over time from registration on the waiting list for a
renal transplant, and to describe the possible differences according to
patients' renal replacement therapy at inclusion in the study,
i.e.~patients either dialyzed or not (preemptive). Indeed, being on the
transplant waiting list may be experienced differently between dialyzed
and preemptive patients. It can be hypothesized that, for example, the
depressive symptoms experienced by dialyzed patients are more \rev{severe}
as compared to preemptive patients due to their experience with dialysis
and expectations of associated complications. It is therefore possible
that the need for clinical and psychological support is not the same for
all patients.

A secondary objective was to assess whether the functioning of some
items differed according to the group (preemptive or dialyzed) or
shifted with time on the waiting list. Indeed, patients may perceive the
items differently according to their renal replacement therapy and over
time, despite having similar depression levels.

\hypertarget{prediala-sample}{%
\subsection{PREDIALA sample}\label{prediala-sample}}

\begin{figure}
\includegraphics[width=0.9\textwidth]{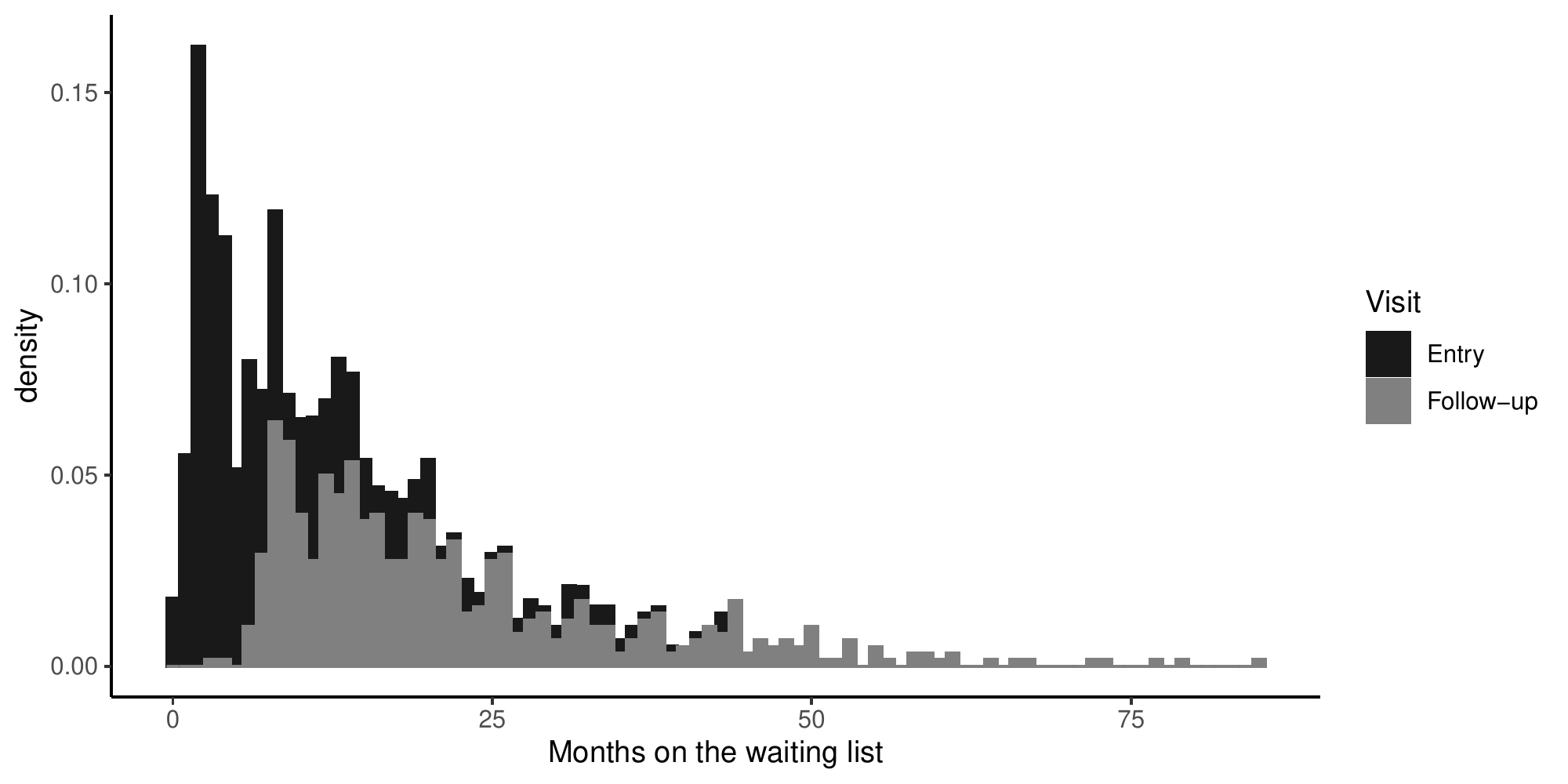}
\caption{Distribution of the measurement times in the PREDIALA study according to the time since registration on the waiting list (at entry in black and during follow-up in grey)}\label{timing}
\end{figure}

We included in the analysis all the patients from PREDIALA who entered
the study within 48 months following their registration on the waiting
list. They were either in the dialyzed or preemptive group at entry, and
had at least one measure for each of the 7 items of the HADS before the
end of the study. The end of the study was defined by either a switch in
group (from preemptive to dialyzed status), a clinical event (mainly
transplantation) or the administrative censoring. From the initial 577 patients included
in PREDIALA study, this selection lead to a final sample of 561 patients
and 1136 repeated visits. Among them 356 (63.5\% ) were men and 288
(51.3\%) were under dialysis. The median age at entry was 59 years
(range 19-67 years), and the patients had been on the waiting list for a
very variable time ranging from 0.1 to 43.1 months (median 5.1 months)
at entry in the cohort. This leads to substantial variability in
measurement timings across patients at entry and during follow-up as
shown in Figure \ref{timing}. This continuous distribution of the
measurement times which would have been ignored using standard IRT
methods is naturally handled in the \rev{longitudinal} IRT model thanks to the
definition of the underlying construct as a latent process in continuous
time.

\hypertarget{the-dynamic-irt-model-specification}{%
\subsection{The \rev{longitudinal} IRT model
specification}\label{the-dynamic-irt-model-specification}}

The \rev{longitudinal} model was defined following equation \eqref{LMM} for the
trajectory of underlying depressive symptomatology and equation
\eqref{probit} for the 7 item-specific measurement model. As we did not
have any assumption regarding the shape of trajectory over time of
depressive symptomatology, we used a basis of natural cubic splines with
2 internal knots placed at tertiles of the measurement time
distribution, that is 7 and 15 months, and boundary knots placed at 0
and 60 months. Each of the four functions of time (intercept and 3
splines functions) was associated with fixed effects specific to the
group (Preemptive/Dialyzed) to assess the mean trajectories, and
individual correlated random effects to account for the correlation
within repeated measures of each individual. For the measurement models,
we assumed in the main analysis that all items functioned similarly,
i.e., no DIF and no response shift occurred. We then explored in
secondary analyses whether some items functioned differently by group
(adding an item-specific contrast on group), and whether some items were
affected by response shift over time (adding an item-specific contrast
of the 3 time functions). The parameter estimates of these three models
are provided in Table S1 of the supplementary material.

\hypertarget{predicted-trajectories-of-depressive-symptomatology}{%
\subsection{Predicted trajectories of depressive
symptomatology}\label{predicted-trajectories-of-depressive-symptomatology}}

\rev{One unit of depressive symptomatology corresponds to the inter-individual variability at
registration in the dialyzed group. In the following description, 1 unit of depressive symptomatology is thus called 1 SD for 1 Standard Deviation}. 

\begin{figure}
\begin{center}
\includegraphics[width=0.8\textwidth]{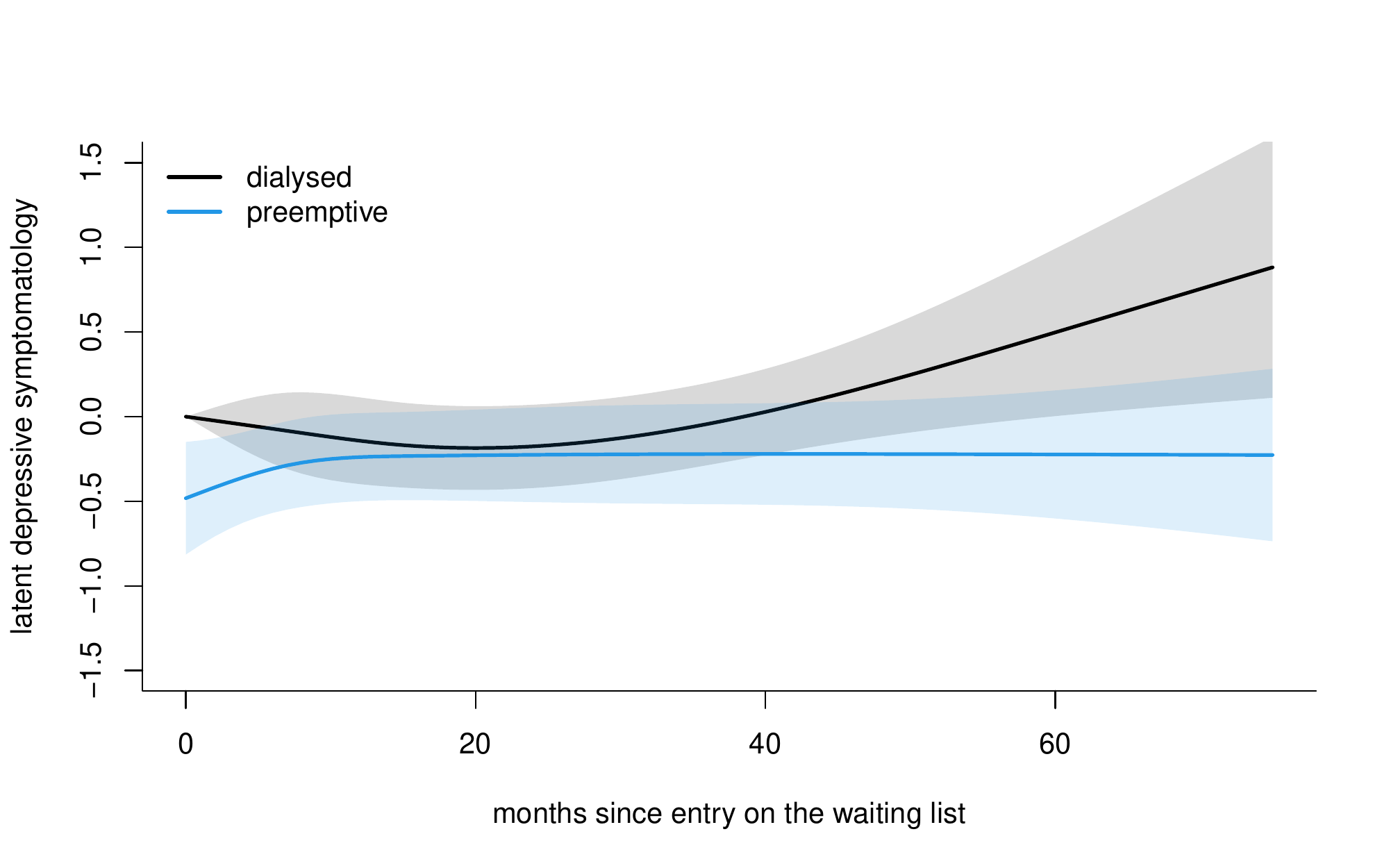}
\end{center}
\caption{Mean trajectory (and 95\% confidence interval in shades) of depressive symptomatology estimated by the \rev{longitudinal} IRT model from the 7 items of HADS repeatedly measured over time; represented for dialyzed and preemptive patients}\label{PredLatent} 
\end{figure}

The \rev{predicted mean} trajectory over time of depressive symptomatology,
displayed in Figure \ref{PredLatent}, varied according to
the group. In the preemptive group, the level of depressive
symptomatology increased during the first year on the waiting list by
0.243 (-0.012,0.498) SD and then remained stable. The level of
depressive symptomatology was higher in the dialyzed group compared to
the preemptive group at the time of registration (difference of -0.482
(-0.814,-0.149) SD). It then slightly decreased during the first year to
reach a similar level as in the preemptive group, and then increased
again after approximately 2 years on the waiting list by a mean annual
rate of 0.245 (0.053,0.438) SD (computed from 2 to 6 years).

\hypertarget{measurement-scale-structure}{%
\subsection{Measurement scale
structure}\label{measurement-scale-structure}}

We exploited the \rev{continuous-time longitudinal} IRT model to assess the HADS depressive
symptomatology items characteristics. To help appreciating the item
characteristics, we assessed the range of the distribution of the
underlying depressive symptomatology based on the estimates of the
model and an hypothetical population of 100000 preemptive
patients and 100000 dialyzed patients with measures every month from
registration up to 72 months. The resulting 95\% prediction interval of
the underlying depressive symptomatology was {[}-6.10,5.90{]} with the
10\% and 90\% percentiles of the distribution at -3.00 and 2.74,
respectively.

\begin{figure}
\begin{center}
\includegraphics[width=0.9\textwidth]{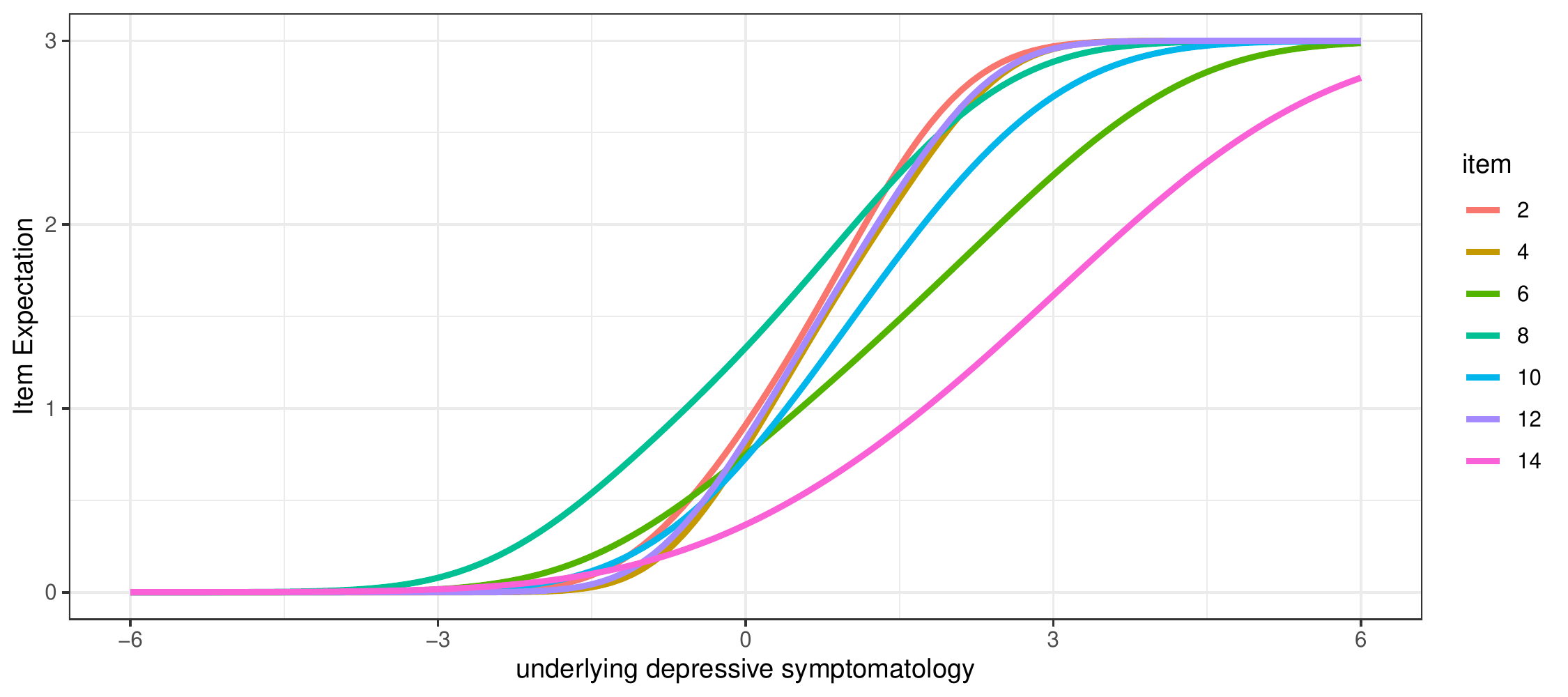}
\includegraphics[width=0.9\textwidth]{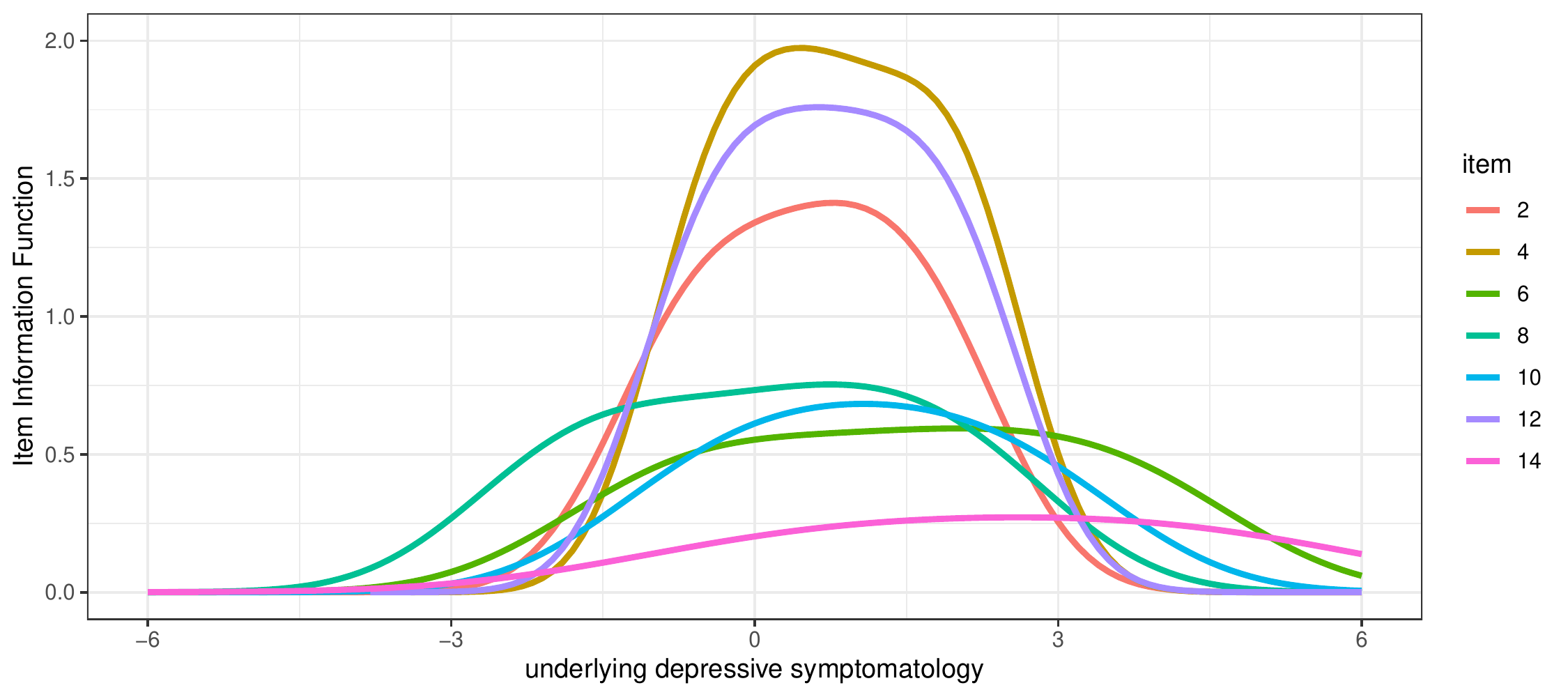}
\end{center}
\caption{Estimated expectation of each item (top) and estimated Fisher information of each item (bottom) according to the underlying depressive symptomatology. Items are rated so that higher levels systematically indicate more intense depressive symptoms}\label{ICC-IIC} 
\end{figure}

Table \ref{LocDisc} provides the estimated locations and discrimination
while Figure \ref{ICC-IIC} shows the curves of item expectations (top)
and curves of item information (bottom) according to the underlying
depressive symptomatology. Figure S1 and S2 of the supplementary
material further display for each item category the probability curve
and the information function, respectively.

The items \rev{with the lowest location along the latent trait} were Item 8
(Slow) and Item 2 (Enjoy) while the \rev{item with the highest location along the latent trait} was Item 14
(Leisure). This means that the level of depression required to respond
to the most unfavorable response categories (i.e.~indicative of \rev{more severe}
depressive symptoms) of items 2 and 8 (e.g., ``Sometimes'' for item 8
``I feel as if I am slowed down'') was lower than the level of
depression required to respond to the most unfavorable response
categories of item 14 (e.g., ``Not often'' for item 14 ``I can enjoy a
good book or radio or TV program''). The most discriminant items with
the steepest curves, representing their ability to discriminate patients
with different levels of depression, were items 2, 4 and 12 concerning
the ability to enjoy, laugh and look forward, respectively. The
estimated curve of the Fisher information plotted in Figure
\ref{ICC-IIC} (bottom) also underlines the major role of these 3 items
compared to the others. In contrast, item 14 about leisure does not
bring much information in this population as it seems to measure much
higher levels of depression than the other items.

\begin{table}[ht]
\centering
\begin{tabular}{lrlrlrlrl}
  \hline
Item  & \multicolumn{2}{c}{category 0 - 1}& \multicolumn{2}{c}{category 1 - 2} & \multicolumn{2}{c}{category 2 - 3} & \multicolumn{2}{c}{discrimination} \\ 
 & est & SE & est & SE & est & SE & est & SE \\ 
  \hline
 2 - Enjoy      & -0.46 & 0.13 & 0.77 & 0.14 & 1.52 & 0.18 & 1.29 & 0.13 \\ 
 4 - Laugh      & -0.26 & 0.12 & 0.74 & 0.13 & 1.91 & 0.21 & 1.56 & 0.16 \\ 
 6 - Cheerful       & -0.48 & 0.14 & 1.58 & 0.20 & 3.34 & 0.38 & 0.85 & 0.09 \\ 
 8 - Slow       & -1.51 & 0.19 & 0.40 & 0.13 & 1.69 & 0.20 & 0.95 & 0.10 \\ 
10 - Appearance     & -0.05 & 0.12 & 1.00 & 0.16 & 2.27 & 0.26 & 0.88 & 0.10 \\ 
12 - Looking Forward    & -0.32 & 0.13 & 0.72 & 0.13 & 1.82 & 0.20 & 1.46 & 0.15 \\ 
14 - Leisure        &  0.83 & 0.17 & 3.18 & 0.42 & 4.11 & 0.54 & 0.56 & 0.07 \\ 
   \hline
\end{tabular}
\caption{Estimate (and associated standard error (SE) obtained by $\Delta$-method) of locations and discrimination of the 7 items of HADS measuring Depressive symptomatology \label{LocDisc}}
\end{table}

\hypertarget{differential-item-functioning-and-response-shift}{%
\subsection{Differential Item Functioning and Response
Shift}\label{differential-item-functioning-and-response-shift}}

We first explored any differential item functioning on the group
(dialyzed versus preemptive). Estimates are provided in Table
\ref{TableDIF} along with those of the main model that ignores DIF.
Overall, the Chi-square test assessing simultaneously the 6 contrasts on
group did not reject the null ``no DIF on group'' assumption (p=0.266).
However, taken individually, the difference between groups for item 2
(Enjoy) was significantly larger than for the other items (item-specific
effect of preemptive group on the underlying level estimated at -0.150
(-0.269,-0.0314) SD). This suggests that this item \rev{has a higher location along the latent trait for
preemptive patients (all location parameters shifted to +0.150) as compared to dialyzed patients for a} same
underlying level of depressive symptomatology; preemptive patients tend
to respond more readily to more favorable response categories than
patients under dialysis, despite having similar depression levels. In
addition, accounting or not for DIF impacted the conclusions: the group
effect on the underlying depressive symptomatology was not significant
anymore when accounting for DIF suggesting that the difference between
groups in the model without DIF was mainly carried by item 2.

\begin{table}[ht]
\centering
\begin{tabular}{rrrrrrrrr}
  \hline
 & & \multicolumn{3}{c}{Without DIF on Group} & &\multicolumn{3}{c}{With DIF on Group} \\
 & & coef & SE & p & & coef & SE & p \\ 
  \hline
intercept & & 0.000 & - & -  & & 0.000 & - & - \\ 
ns1 & &-0.305 & 0.172 & 0.075 & & -0.304 & 0.229 & 0.184 \\ 
ns2 & &0.039 & 0.222 & 0.862 & & 0.038 & 0.590 & 0.949 \\ 
ns3 & &0.538 & 0.250 & 0.031 & & 0.538 & 0.310 & 0.083 \\ 
  group& & -0.482 & 0.170 & 0.004 & & -0.458 & 0.413 & 0.267 \\ 
ns1:preemptive& & 0.477 & 0.223 & 0.032&  & 0.473 & 0.265 & 0.075 \\ 
ns2:preemptive& & 0.428 & 0.335 & 0.201 & & 0.429 & 0.491 & 0.382 \\ 
ns3:preemptive& & -0.390 & 0.292 & 0.182 & & -0.391 & 0.338 & 0.247 \\ 
\emph{Contrasts on preemptive:} & &  &  & &  & \multicolumn{3}{c}{\emph{(global p=0.266)} } \\ 
Item 2 &&  &  &  & & -0.150 & 0.061 & 0.013 \\ 
Item 4 &&  &  &  & & -0.031 & 0.059 & 0.600 \\ 
Item 6 &&  &  &  & & 0.020 & 0.077 & 0.800 \\ 
Item 8 &&  &  &  & & -0.053 & 0.100 & 0.593 \\ 
Item 10 &&  &  &  & & 0.040 & 0.158 & 0.802 \\ 
Item 12 &&  &  &  & & 0.037 & 0.061 & 0.543 \\ 
** Item 14&&  &  &  & & 0.138 & 0.124 & 0.266 \\ 
   \hline
\end{tabular}

 ** coefficient not estimated but obtained as minus the sum of the others.
\caption{Estimated fixed parameters in the \rev{longitudinal IRT model} without (left) and with (right) differential item functioning (DIF) on group. ns1, ns2, ns3 refer to the natural cubic splines functions.}\label{TableDIF}
\end{table}

\begin{figure} 
\includegraphics[width=1\textwidth]{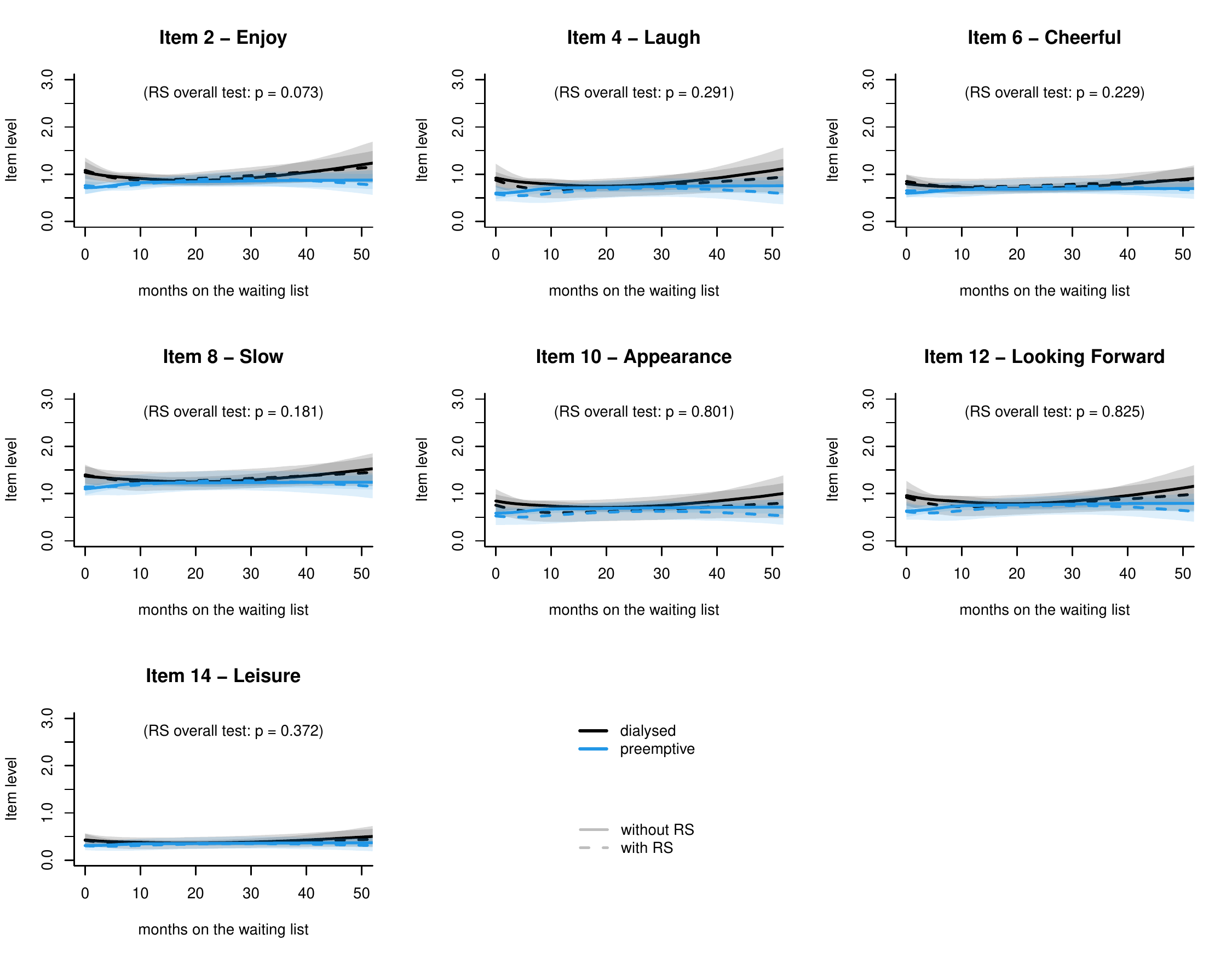}
\caption{Predicted items trajectories according to time on the waiting list and group in the model neglecting potential response shift (plain) and in the one accounting for potential response shift (dashed) - with p-value of the overall \rev{item-specific Wald} test for RS over the 3 splines functions in the model assuming RS.}\label{predRS}
\end{figure}

We secondly explored item response shift over time by adding
item-specific contrasts on the 3 natural cubic splines functions of time
(ns1, ns2, ns3). \rev{This resulted in 18 additional parameters (6 contrasts on 3 time functions). Estimates are given in supplementary Table S2. The likelihood ratio test for these 18 parameters, which provides an overall test for RS, did not reject the null hypothesis with a statistic of 22.9 points (p=0.194). The RS was further assessed item by item by testing the 3 item-specific contrasts on time in a multivariate Wald test. Again, no evidence of RS was found for all the items (p$>$0.181) except maybe for item 2 (p=0.073).} The predicted trajectories of each item (Figure \ref{predRS}) in the models with and without RS confirmed the absence of substantial response shift over time. Each item behavior over time was close
to the one of the underlying construct showed in Figure \ref{PredLatent}
despite some slight differences in the intensity of change in the first
months after entry on the waiting list. 

\hypertarget{conclusions}{%
\section{Conclusions}\label{conclusions}}

We have described how to combine the item response theory with the
linear mixed model theory for item-level analysis of longitudinal PRO or
CRO data when measurement times may vary across individuals. Using a
real case example with the PREDIALA study, we have shown that this
\rev{longitudinal} IRT model can describe the latent construct trajectory over time
and its determinants, while simultaneously assessing the item and scale
properties, and exploring lack of measurement invariance between groups
(DIF) and over time (response shift).

This analysis of PREDIALA data helped to better understand the experience of patients with end-stage renal
disease on the renal transplant waiting list in terms of depressive
symptoms. DIF was highlighted on item 2 (Enjoy) indicating that patients
under dialysis had more difficulty in reporting having enjoyment than
preemptive patients, despite having similar levels of depression. This
may reveal that the need for clinical and psychological support may not
be the same for all patients, according to their renal replacement
therapy. Response shift was not significantly evidenced despite a trend
on this same item 2. Adjusting for DIF and response shift, the level of
depressive symptomatology of preemptive patients tended to slightly
increase during the first year and to remain stable afterwards. The
level of depressive symptomatology of patients under dialysis tended to
be close to the one of preemptive patients after 2 years on the waiting
list and to increase afterwards. Although it has been reported that the
time on waiting list should be reduced to limit depressive symptoms
\cite{corruble2010} and improve health-related quality of life \cite{ong2013}, the shortage of grafts unfortunately often makes this
difficult to achieve.\\

The \rev{longitudinal} IRT model we described here unites the strengths of IRT and LMM theories. On
the one hand, the use of a structural mixed model makes it possible to
operationalize the latent construct as a latent process defined in
continuous time and thus takes into account that most health phenomena
intrinsically evolve in continuous time. On the other hand, the use of
IRT methodology to define the measurement scale at each individual- and
occasion-specific observation time enables a precise modeling of the
items constituting the measurement scale, and their properties. \rev{Longitudinal IRT models had already been proposed in the literature. However, following the usual specification of latent variable models, most of them had considered time as discrete by defining a different latent variable at each measurement occasion (although all these latent variables are supposed to measure the same construct) \cite{wang2020,cai2021}. In addition to not \rev{aligning} with the continuous-time nature of health phenomena, it prevents their use in \rev{some} applications, such as in our study, where measurement times largely vary across subjects. Moreover, neglecting the different lengths of time intervals in longitudinal IRT models \rev{may} provide biased estimates \cite{hecht2019}.
The issue of continuous time in longitudinal psychological studies had already been identified in the literature \cite{voelkle2018} and dealt with by considering continuous-time dynamic systems (with differential equations) to capture the intra-individual serial correlation \cite{hecht2019} and implemented in ctsem R package \cite{driver2017}. With the use of a structural mixed model rather than a structural dynamic model, we offer with lcmm R package an alternative approach to flexibly model latent traits in continuous time. We note that, although commercial Mplus software \cite{muthen1998} was initially dedicated to discrete-time analyses, it now also offers the posibility to consider time as continuous when defining simple shapes of trajectory (e.g., linear or quadratic), or time as finely discretized with autoregressive structures \citep{asparouhov2018}. The definition of the underlying construct as a latent process evolving in continuous time enables the inclusion of items whatever their measurement timing. These timings can differ from one subject to the other (as shown in the application) and they can differ from one item to the other. We recommend however that the span of measurement times is not too dissimilar, at least for a subset of the items, to ensure that the latent process can be correctly anchored during the entire study time window}.  \\

As for all methodologies, the method we propose is not without some
limitations and some further path of research may be put forward. First,
we focused here on a specific measurement model, the GRM, which
translates the discretization of the underlying latent process into
ordinal categories as shown in Equation \eqref{explic} \cite{commenges2015}. However, coming from IRT models, this measurement model does not
possess the specific objectivity property as Rasch Measurement Theory
(RMT) models do \cite{andrich2011}. It would thus be interesting to adapt
the proposed methodology to RMT measurement models (see Blanchin et
al. \citep{blanchin2021} in this special issue and Barbieri et al. \cite{barbieri2017a}). Of note, changing
the measurement model does not impact the estimation procedure nor the
structural part of the model, and the current version of the program
already handles different measurement models for continuous items in
addition to GRM for ordinal items \rev{with the possibility to mix continuous and categorical items within the same model}. 

\rev{Second, in the current specification, the serial intra-individual correlation is fully captured by the random effects. Since any type and dimension of time functions can be entered in the model, the method can handle a large variety of structures of intra-individual correlation, whether the data are spaced or frequent. However, in some contexts, considering an additional Gaussian autocorrelation process may be of interest. Although not considered in our work, this could be easily added to the structural linear mixed model as it was previously done with continuous outcomes \citep{proust2006,proust-lima2013}. Indeed, the added numerical complexity would be low given the use of the quasi Monte-Carlo integral approximation in the log-likelihood which does not vary much with the dimensionality of the random deviations}.  

\rev{Third,} by relying on the mixed model theory and the maximum likelihood
estimation, the \rev{longitudinal} IRT model relies on the missing at random
assumption both for monotonic and for intermittent missingness \cite{little1995}. In the presence of informative dropout, a joint model for the
repeated item responses and the time to dropout, as described by
Saulnier et al.~ \citep{saulnier2021} in this special issue, should be favored. This
joint model combines the \rev{longitudinal} IRT model with a survival model that
captures the association between the underlying latent construct and the
dropout (or any other event of interest). 

Fourth, the methodology is
fully parametric, so that analytic choices are systematically made. For
instance, to simplify the application setting, we only included
time-independent covariates even though time-dependent covariates could
also be considered, such as a changing of group during the follow-up. We
also globally tested the lack of measurement invariance over the items'
parameters across the overall follow-up although a more precise
assessment could be done regarding which function of time is affected or
at which time the lack of measurement invariance occurs.\\

To conclude, by extending the IRT methodology to longitudinal data, and
considering the time as continuous, our methodology provides a versatile
and flexible approach for modeling item responses measured repeatedly
over time as encountered in numerous longitudinal health studies.

\newpage

\textbf{Acknowledgements} The authors gratefully thank the
co-investigators of the study: Magali Giral, Aurélie Meurette, Emmanuel
Morelon, and Laetitia Albano. The authors express sincere thanks to
Elodie Faurel-Paul and Astrid Fleury for their assistance for the study,
as well as all the participants for their contribution to the study. The
authors wish to thank members of the clinical research assistant team
and DIVAT Consortium Collaborators (Medical Doctors, Surgeons, HLA
Biologists). Nantes: Gilles Blancho, Julien Branchereau, Diego
Cantarovich, Agnès Chapelet, Jacques Dantal, Clément Deltombe, Lucile
Figueres, Claire Garandeau, Magali Giral, Caroline Gourraud-Vercel,
Maryvonne Hourmant, Georges Karam, Clarisse Kerleau, Aurélie Meurette,
Simon Ville, Christine Kandell, Anne Moreau, Karine Renaudin, Anne
Cesbron, Florent Delbos, Alexandre Walencik, Anne Devis; Lyon E. Hériot
: Lionel Badet, Maria Brunet, Fanny Buron, Rémi Cahen, Sameh Daoud,
Coralie Fournie, Arnaud Grégoire, Alice Koenig, Charlène Lévi, Emmanuel
Morelon, Claire Pouteil-Noble, Thomas Rimmelé, Olivier Thaunat.

\textbf{Funding} This work was funded by the French National Research
Agency (Project DyMES - ANR-18-C36-0004-01), the French Ministry of
Health (PHRC-13-0224, 2013) and by the "Appel d offres interne analyse secondaire 2018" grant from Nantes University Hospital. 

\textbf{Ethical approval} The PREDIALA study is part of the PreKit-QoL
study which is registered on the ClinicalTrials.gov Registry
(RC14\_0078, NCT02154815). It has obtained approval from the ethical
Committee for Persons' Protection (CPP, Tours, 2014-S8), and from the
advisory committee on research data and information in health (CCTIRS,
Paris, 14.314).

\textbf{Conflict of interest} All authors have no competing interest.

\textbf{Supplementary Material} Supplementary material can be found on
the journal website.

\textbf{Data availability} The software is openly available in the R
package lcmm (on \url{https://github.com/CecileProust-Lima/lcmm} and on
cran). The R script along with a dataset mimicking the PREDIALA study
are also provided as a package vignette. The raw PREDIALA data can not
be shared.

\bibliography{dynamicIRT_submission_HAL_revised}

\begin{thebibliography}{10}
\expandafter\ifx\csname url\endcsname\relax
  \def\url#1{\texttt{#1}}\fi
\expandafter\ifx\csname urlprefix\endcsname\relax\def\urlprefix{URL }\fi
\expandafter\ifx\csname href\endcsname\relax
  \def\href#1#2{#2} \def\path#1{#1}\fi

\bibitem{cerou2019}
M.~{Cerou}, S.~{Peigné}, E.~{Comets}, M.~{Chenel}, Application of item
  response theory to model disease progression and agomelatine effect in
  patients with major depressive disorder, The AAPS journal 22~(1) (2019) 4,
  pMID: 31720897.
\newblock \href {http://dx.doi.org/10.1208/s12248-019-0379-x}
  {\path{doi:10.1208/s12248-019-0379-x}}.

\bibitem{gorter2015}
R.~Gorter, J.-P. Fox, J.~W.~R. Twisk,
  \href{https://doi.org/10.1186/s12874-015-0050-x}{Why item response theory
  should be used for longitudinal questionnaire data analysis in medical
  research}, BMC Medical Research Methodology 15~(1) (2015) 55.
\newblock \href {http://dx.doi.org/10.1186/s12874-015-0050-x}
  {\path{doi:10.1186/s12874-015-0050-x}}.
\newline\urlprefix\url{https://doi.org/10.1186/s12874-015-0050-x}

\bibitem{mccall2021}
W.~{McCall}, B.~{Porter}, A.~{Pate}, C.~{Bolstad}, C.~{Drapeau}, A.~{Krystal},
  R.~{Benca}, M.~{Rumble}, M.~{Nadorff}, Examining suicide assessment measures
  for research use: Using item response theory to optimize psychometric
  assessment for research on suicidal ideation in major depressive disorder,
  Suicide \& Life-Threatening BehaviorPMID: 34237156.
\newblock \href {http://dx.doi.org/10.1111/sltb.12791}
  {\path{doi:10.1111/sltb.12791}}.

\bibitem{abdelhamid2021}
G.~{Abdelhamid}, M.~{Bassiouni}, J.~{Gómez-Benito}, Assessing cognitive
  abilities using the wais-iv: An item response theory approach, International
  Journal of Environmental Research and Public Health 18~(13) (2021) 6835,
  pMID: 34202249 PMCID: PMC8297006.
\newblock \href {http://dx.doi.org/10.3390/ijerph18136835}
  {\path{doi:10.3390/ijerph18136835}}.

\bibitem{rakers2021}
S.~E. Rakers, M.~E. Timmerman, M.~E. Scheenen, M.~E. de~Koning, H.~J. van~der
  Horn, J.~van~der Naalt, J.~M. Spikman, Trajectories of {Fatigue},
  {Psychological} {Distress}, and {Coping} {Styles} {After} {Mild} {Traumatic}
  {Brain} {Injury}: {A} 6-{Month} {Prospective} {Cohort} {Study}, Archives of
  Physical Medicine and Rehabilitation (2021) S0003--9993(21)00462--7\href
  {http://dx.doi.org/10.1016/j.apmr.2021.06.004}
  {\path{doi:10.1016/j.apmr.2021.06.004}}.

\bibitem{otto2021}
I.~Otto, C.~Hilger, A.~Magheli, G.~Stadler, F.~Kendel, Illness representations,
  coping and anxiety among men with localized prostate cancer over an
  18-months period: {A} parallel vs. level-contrast mediation approach,
  Psycho-Oncology\href {http://dx.doi.org/10.1002/pon.5798}
  {\path{doi:10.1002/pon.5798}}.

\bibitem{edjolo2016}
A.~Edjolo, C.~Proust-Lima, F.~Delva, J.-F. Dartigues, K.~Pérès, Natural
  {History} of {Dependency} in the {Elderly}: {A} 24-{Year}
  {Population}-{Based} {Study} {Using} a {Longitudinal} {Item} {Response}
  {Theory} {Model}, American Journal of Epidemiology 183~(4) (2016) 277--285,
  number: 4.
\newblock \href {http://dx.doi.org/10.1093/aje/kwv223}
  {\path{doi:10.1093/aje/kwv223}}.

\bibitem{wang2020}
C.~Wang, S.~W. Nydick, \href{https://doi.org/10.3102/1076998619882026}{On
  {Longitudinal} {Item} {Response} {Theory} {Models}: {A} {Didactic}}, Journal
  of Educational and Behavioral Statistics 45~(3) (2020) 339--368, publisher:
  American Educational Research Association.
\newblock \href {http://dx.doi.org/10.3102/1076998619882026}
  {\path{doi:10.3102/1076998619882026}}.
\newline\urlprefix\url{https://doi.org/10.3102/1076998619882026}

\bibitem{cai2021}
L.~Cai, C.~R. Houts, Longitudinal {Analysis} of {Patient}-{Reported} {Outcomes}
  in {Clinical} {Trials}: {Applications} of {Multilevel} and {Multidimensional}
  {Item} {Response} {Theory}, Psychometrika 86~(3) (2021) 754--777.
\newblock \href {http://dx.doi.org/10.1007/s11336-021-09777-y}
  {\path{doi:10.1007/s11336-021-09777-y}}.

\bibitem{laird1982}
N.~M. Laird, J.~H. Ware, Random-effects models for longitudinal data,
  Biometrics 38~(4) (1982) 963--74.

\bibitem{samejima1997}
F.~Samejima, \href{https://doi.org/10.1007/978-1-4757-2691-6_5}{Graded
  {Response} {Model}}, in: W.~J. van~der Linden, R.~K. Hambleton (Eds.),
  Handbook of {Modern} {Item} {Response} {Theory}, Springer, New York, NY,
  1997, pp. 85--100.
\newblock \href {http://dx.doi.org/10.1007/978-1-4757-2691-6_5}
  {\path{doi:10.1007/978-1-4757-2691-6_5}}.
\newline\urlprefix\url{https://doi.org/10.1007/978-1-4757-2691-6_5}

\bibitem{sebille2016}
V.~Sébille, J.-B. Hardouin, M.~Giral, A.~Bonnaud-Antignac, P.~Tessier,
  E.~Papuchon, A.~Jobert, E.~Faurel-Paul, S.~Gentile, E.~Cassuto, E.~Morélon,
  L.~Rostaing, D.~Glotz, R.~Sberro-Soussan, Y.~Foucher, A.~Meurette,
  Prospective, multicenter, controlled study of quality of life, psychological
  adjustment process and medical outcomes of patients receiving a preemptive
  kidney transplant compared to a similar population of recipients after a
  dialysis period of less than three years--{The} {PreKit}-{QoL} study
  protocol, BMC nephrology 17 (2016) 11.
\newblock \href {http://dx.doi.org/10.1186/s12882-016-0225-7}
  {\path{doi:10.1186/s12882-016-0225-7}}.

\bibitem{auneau-enjalbert2020}
L.~Auneau-Enjalbert, J.-B. Hardouin, M.~Blanchin, M.~Giral, E.~Morelon,
  E.~Cassuto, A.~Meurette, V.~Sébille, Comparison of longitudinal quality of
  life outcomes in preemptive and dialyzed patients on waiting list for kidney
  transplantation, Quality of Life Research: An International Journal of
  Quality of Life Aspects of Treatment, Care and Rehabilitation 29~(4) (2020)
  959--970.
\newblock \href {http://dx.doi.org/10.1007/s11136-019-02372-w}
  {\path{doi:10.1007/s11136-019-02372-w}}.

\bibitem{tong2015}
A.~Tong, C.~S. Hanson, J.~R. Chapman, F.~Halleck, K.~Budde, M.~A. Josephson,
  J.~C. Craig, '{Suspended} in a paradox'-patient attitudes to wait-listing for
  kidney transplantation: systematic review and thematic synthesis of
  qualitative studies, Transplant International: Official Journal of the
  European Society for Organ Transplantation 28~(7) (2015) 771--787.
\newblock \href {http://dx.doi.org/10.1111/tri.12575}
  {\path{doi:10.1111/tri.12575}}.

\bibitem{holland2009}
P.~W. Holland, H.~Wainer (Eds.), Differential {Item} {Functioning}, Routledge,
  New-York, NY, 2009.

\bibitem{sprangers1999}
M.~A. Sprangers, C.~E. Schwartz, Integrating response shift into health-related
  quality of life research: a theoretical model, Social Science \& Medicine
  (1982) 48~(11) (1999) 1507--1515.
\newblock \href {http://dx.doi.org/10.1016/s0277-9536(99)00045-3}
  {\path{doi:10.1016/s0277-9536(99)00045-3}}.

\bibitem{proust2006}
C.~Proust, H.~Jacqmin-Gadda, J.~M.~G. Taylor, J.~Ganiayre, D.~Commenges, A
  nonlinear model with latent process for cognitive evolution using
  multivariate longitudinal data, Biometrics 62~(4) (2006) 1014--1024.
\newblock \href {http://dx.doi.org/10.1111/j.1541-0420.2006.00573.x}
  {\path{doi:10.1111/j.1541-0420.2006.00573.x}}.

\bibitem{proust-lima2013}
C.~Proust-Lima, H.~Amieva, H.~Jacqmin-Gadda, Analysis of multivariate mixed
  longitudinal data: a flexible latent process approach, The British journal of
  mathematical and statistical psychology 66~(3) (2013) 470--487.
\newblock \href {http://dx.doi.org/10.1111/bmsp.12000}
  {\path{doi:10.1111/bmsp.12000}}.

\bibitem{baker2004}
F.~B. Baker, S.~H. Kim, Item {Response} {Theory}. {Parameter} {Estimation}
  {Techniques}, 2nd Edition, Statistics: {Textbooks} \& {Monographs}, Marcel
  Dekker, New York, 2004.

\bibitem{saulnier2021}
T.~Saulnier, V.~Philipps, W.~G. Meissner, O.~Rascol, A.~Pavy-Le-Traon,
  A.~Foubert-Samier, C.~Proust-Lima,
  \href{http://arxiv.org/abs/2110.02612}{Joint models for the longitudinal
  analysis of measurement scales in the presence of informative dropout},
  arXiv:2110.02612 [stat]ArXiv: 2110.02612.
\newline\urlprefix\url{http://arxiv.org/abs/2110.02612}

\bibitem{barbieri2017a}
A.~Barbieri, J.~Peyhardi, T.~Conroy, S.~Gourgou, C.~Lavergne, C.~Mollevi,
  \href{https://doi.org/10.1186/s12874-017-0410-9}{Item response models for the
  longitudinal analysis of health-related quality of life in cancer clinical
  trials}, BMC Medical Research Methodology 17~(1) (2017) 148.
\newblock \href {http://dx.doi.org/10.1186/s12874-017-0410-9}
  {\path{doi:10.1186/s12874-017-0410-9}}.
\newline\urlprefix\url{https://doi.org/10.1186/s12874-017-0410-9}

\bibitem{philipson2020}
P.~Philipson, G.~L. Hickey, M.~J. Crowther, R.~Kolamunnage-Dona,
  \href{https://www.sciencedirect.com/science/article/pii/S0167947320301018}{Faster
  {Monte} {Carlo} estimation of joint models for time-to-event and multivariate
  longitudinal data}, Computational Statistics \& Data Analysis 151 (2020)
  107010.
\newblock \href {http://dx.doi.org/10.1016/j.csda.2020.107010}
  {\path{doi:10.1016/j.csda.2020.107010}}.
\newline\urlprefix\url{https://www.sciencedirect.com/science/article/pii/S0167947320301018}

\bibitem{philipps2021}
V.~Philipps, B.~P. Hejblum, M.~Prague, D.~Commenges, C.~Proust-Lima, Robust and
  efficient optimization using a {Marquardt}-{Levenberg} algorithm with {R}
  package {marqLevAlg}, R journal in press.

\bibitem{proust-lima2017}
C.~Proust-Lima, V.~Philipps, B.~Liquet,
  \href{https://www.jstatsoft.org/v078/i02}{Estimation of {Extended} {Mixed}
  {Models} {Using} {Latent} {Classes} and {Latent} {Processes}: {The} {R}
  {Package} lcmm}, Journal of Statistical Software, Articles 78~(2) (2017)
  1--56.
\newblock \href {http://dx.doi.org/10.18637/jss.v078.i02}
  {\path{doi:10.18637/jss.v078.i02}}.
\newline\urlprefix\url{https://www.jstatsoft.org/v078/i02}

\bibitem{zigmond1983}
A.~S. Zigmond, R.~P. Snaith, The hospital anxiety and depression scale, Acta
  Psychiatrica Scandinavica 67~(6) (1983) 361--370.
\newblock \href {http://dx.doi.org/10.1111/j.1600-0447.1983.tb09716.x}
  {\path{doi:10.1111/j.1600-0447.1983.tb09716.x}}.

\bibitem{corruble2010}
E.~Corruble, A.~Durrbach, B.~Charpentier, P.~Lang, S.~Amidi, A.~Dezamis,
  C.~Barry, B.~Falissard,
  \href{https://doi.org/10.1080/08964280903521339}{Progressive {Increase} of
  {Anxiety} and {Depression} in {Patients} {Waiting} for a {Kidney}
  {Transplantation}}, Behavioral Medicine 36~(1) (2010) 32--36, publisher:
  Taylor \& Francis \_eprint: https://doi.org/10.1080/08964280903521339.
\newblock \href {http://dx.doi.org/10.1080/08964280903521339}
  {\path{doi:10.1080/08964280903521339}}.
\newline\urlprefix\url{https://doi.org/10.1080/08964280903521339}

\bibitem{ong2013}
S.~C. Ong, W.~L. Chow, S.~van~der Erf, V.~D. Joshi, J.~F. Lim, C.~Lim, P.~S.
  Tee, Y.~M. Lu, T.~Y. Kee, What factors really matter? {Health}-related
  quality of life for patients on kidney transplant waiting list, Annals of the
  Academy of Medicine, Singapore 42~(12) (2013) 657--666.

\bibitem{hecht2019}
M.~Hecht, K.~Hardt, C.~C. Driver, M.~C. Voelkle, Bayesian continuous-time
  {Rasch} models, Psychological Methods 24~(4) (2019) 516--537, place: US
  Publisher: American Psychological Association.
\newblock \href {http://dx.doi.org/10.1037/met0000205}
  {\path{doi:10.1037/met0000205}}.

\bibitem{voelkle2018}
M.~C. Voelkle, C.~Gische, C.~C. Driver, U.~Lindenberger, The {Role} of {Time}
  in the {Quest} for {Understanding} {Psychological} {Mechanisms}, Multivariate
  Behavioral Research 53~(6) (2018) 782--805, number: 6.
\newblock \href {http://dx.doi.org/10.1080/00273171.2018.1496813}
  {\path{doi:10.1080/00273171.2018.1496813}}.

\bibitem{driver2017}
C.~C. Driver, J.~H.~L. Oud, M.~C. Voelkle,
  \href{https://doi.org/10.18637/jss.v077.i05}{Continuous {Time} {Structural}
  {Equation} {Modeling} with {R} {Package} ctsem}, Journal of Statistical
  Software 77 (2017) 1--35.
\newblock \href {http://dx.doi.org/10.18637/jss.v077.i05}
  {\path{doi:10.18637/jss.v077.i05}}.
\newline\urlprefix\url{https://doi.org/10.18637/jss.v077.i05}

\bibitem{muthen1998}
B.~O. Muthén, L.~K. Muthén, Mplus [computer software] (1998-2021).

\bibitem{asparouhov2018}
T.~Asparouhov, E.~L. Hamaker, B.~Muthén,
  \href{https://doi.org/10.1080/10705511.2017.1406803}{Dynamic structural
  equation models}, Structural Equation Modeling: A Multidisciplinary Journal
  25~(3) (2018) 359--388.
\newblock \href
  {http://arxiv.org/abs/https://doi.org/10.1080/10705511.2017.1406803}
  {\path{arXiv:https://doi.org/10.1080/10705511.2017.1406803}}, \href
  {http://dx.doi.org/10.1080/10705511.2017.1406803}
  {\path{doi:10.1080/10705511.2017.1406803}}.
\newline\urlprefix\url{https://doi.org/10.1080/10705511.2017.1406803}

\bibitem{commenges2015}
D.~Commenges, H.~Jacqmin-Gadda, A.~Alioum, P.~Joly, B.~Liquet, C.~Proust-Lima,
  V.~Rondeau, R.~Thiébaut, Chapter 5. {Extensions} of mixed models, in:
  Dynamical {Biostastical} {Models}, CRC Press, 2015.

\bibitem{andrich2011}
D.~{Andrich}, Rating scales and rasch measurement, Expert Review of
  Pharmacoeconomics \& Outcomes Research 11~(5) (2011) 571--585, pMID:
  21958102.
\newblock \href {http://dx.doi.org/10.1586/erp.11.59}
  {\path{doi:10.1586/erp.11.59}}.

\bibitem{blanchin2021}
M.~Blanchin, b.~priscilla, V.~S{\'e}bille,
  \href{https://hal.archives-ouvertes.fr/hal-03389050}{{Performance of a
  Rasch-based method for group comparisons of longitudinal change and response
  shift at item-level in PRO data: a simulation study}}, working paper or
  preprint (Sep. 2021).
\newline\urlprefix\url{https://hal.archives-ouvertes.fr/hal-03389050}

\bibitem{little1995}
R.~J.~A. Little,
  \href{http://amstat.tandfonline.com/doi/abs/10.1080/01621459.1995.10476615}{Modeling
  the {Drop}-{Out} {Mechanism} in {Repeated}-{Measures} {Studies}}, Journal of
  the American Statistical Association 90~(431) (1995) 1112--1121.
\newblock \href {http://dx.doi.org/10.1080/01621459.1995.10476615}
  {\path{doi:10.1080/01621459.1995.10476615}}.
\newline\urlprefix\url{http://amstat.tandfonline.com/doi/abs/10.1080/01621459.1995.10476615}

\end{thebibliography}

\end{document}